# Inclusive Design: Accessibility Settings for People with Cognitive Disabilities

Trae Waggoner, Julia Ann Jose, Ashwin Nair, and Sudarsan Manikandan
Arizona State University, tlwaggon@asu.edu, jajose2@asu.edu, agnair@asu.edu, smanika1@asu.edu

*Abstract -* The advancement of technology has progressed faster than any other field in the world. And with the development of these new technologies, it is important to make sure that these tools can be used by everyone, including people with disabilities. Accessibility options in computing devices help ensure that everyone has the same access to advanced technologies. Unfortunately, for those who require more unique and sometimes challenging accommodations, such as people with Amyotrophic lateral sclerosis (ALS), the most commonly used accessibility features are simply not enough. While assistive technology for those with ALS does exist, it requires multiple peripheral devices that can become quite expensive collectively. The purpose of this paper is to suggest a more affordable and readily available option for ALS assistive technology that can be implemented on a smartphone or tablet.

## 1. INTRODUCTION

People who have ALS often lose the ability to speak, swallow, or breathe on their own because of the disease's connection to muscle control. Curative therapies for people with ALS do not currently exist. Accordingly, maintaining a person's quality of life becomes the most important factor for people with ALS. Having access to advanced technologies significantly improves quality of life, especially for people with disabilities. In this paper, we propose an accessibility setting that helps people with speech and motor difficulties such as ALS patients use smartphones using their eyes.

### 1.1. SOLUTION OVERVIEW

For people with speech and motor difficulties, the best accessibility option is offered by eye tracking technology. The proposed system uses cameras to track the position of the user's eyes relative to the position of on-screen images. The system then observes the user's blinks to select certain objects on-screen for the user to manipulate. Using this sequence of eye tracking and blinking, the user can perform various actions such as typing a message, selecting applications, playing games, etc.

While other companies have developed eye tracking systems, they often require a bevy of devices and peripherals to operate. When added together, these systems can become quite expensive. Additionally, these systems often require the user to pair them to their phone or computer in order for the technology to be assistive. Our proposed solution is to integrate this eye tracking technology directly onto the user's smartphone. Using the peripherals already built into most smartphones, this eye tracking technology can perform the tasks of the current iterations while saving on expenses and resources.

### 1.2. RELEVANCE TO MOBILE COMPUTING

The proposed solution will be integrated into a smartphone or tablet. The system will utilize the peripherals already built into these devices such as a built-in IR camera. To this end, it must reliably interact with these components as well as the mobile applications on the device. Additionally, the system will have to balance power consumption and time/space complexity with reliability and usability.

## 2. RELATED WORKS

There are many assistive technology devices available for people with motor and speech difficulties. These devices help people with conditions such as ALS, to move around, communicate with others, etc. There's a plethora of speech generating devices available such as [2]. These devices rely on the patient's caregiver for further support. [3] is yet another powerful eye-driven tablet for communication purposes. As shown in Figure 1, This device has a camera mounted below it's screen that tracks the user's eye movement and helps the user communicate with the device using their eyes.

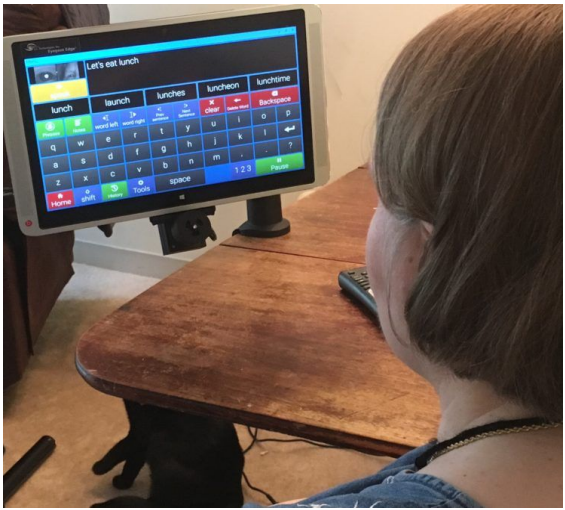

*Figure 1: Eyegaze Edge*

While many such devices exist, our primary interest behind this paper is on how we can make this an inclusive accessibility setting on our mobile phones and on how we can improve eye-tracking and gaze estimation to make it background invariant, lighting invariant and head-pose invariant, thereby improving some of the limitations and challenges of some of these existing technologies. We focus on how we can deploy this functionality as an accessibility setting on our smartphone devices without having the need to use external sensors, equipment or even external wearables.

The problem of eye-tracking is an extensive research topic and there exists several proposed algorithms for this. These can be mainly classified into commercial eye-gaze devices and machine learning based eye detectors. Under the commercial eye-gaze devices category, several mathematical and optical models have been studied and have been used to analyse eye movements. [4] uses an eye-center localization algorithm and geometrical models to gain insight into how eye-movements can be captured. [5] explains an eye-center localization technique by means of image gradients. Under the Machine learning based eye-detector category, we have several interesting algorithms too. [6] talks about how discriminatory Haar features and an SVM model could be used for feature extraction and classification purposes. These techniques still face the challenge of accuracy under low resolution conditions, low/high contrast conditions, lighting variations etc. [7] attempts to solve these problems using Convolutional Neural Networks. They make use of a very large dataset obtained by means of crowdsourcing using Amazon Mechanical Turk platform. Because of the model complexity, the model becomes computationally expensive to use on mobile devices. They solve this problem by using the concept of dark knowledge [8] while compromising on accuracy and convergence time to some extent.

## 3. PROPOSED SOLUTION

Our proposed solution is to have an additional mode in the smartphone's accessibility setting called *i-screen*. When enabled, the phone switches into a mode — one that is eye-controlled.

Under this mode, one can use their eyes to control their phone. To tap on app icons, the user would blink. To type, the user would blink at the first letter and then gaze through the rest of the letters. The predictive text and auto-correct features would still be available and the user would blink to select it. To scroll up, the user would move their eyes from bottom to top and to scroll down, the user would move their eyes from top to bottom.

Our proposed solution to implement this mode can be divided into three tasks — eye-detection, collection of training data and training a machine learning model for eye-tracking.

1. *Eye Detection*
   In this task, we propose to use a built-in (smartphone) IR (infra-red) based camera for eye-tracking.

We propose that a camera that detects light in near-infrared spectrum light be embedded into these smartphones. When our eyes are exposed to near-infrared light, it causes a certain type of reflection called Pupil Center Corneal Reflection [9]. As a result, our eyes show two bright glints (reflections) in both the pupil and the cornea. This image of our eye can then be captured by the camera and a vector can be formed using the pupil and corneal deflections and this vector along with other mathematical techniques can be used to estimate our gaze. [9,10,12,13]

We also aim to associate a 'no reflection' image to a 'blink'. If a blink is encountered, this would indicate that the user wants to click/tap at that point which the user was looking at right before blinking. The blink feature can be used to tap on the app icons on the smartphone's screen or even to select or click on buttons.

In this way, narrowing down the dataset for the machine learning model to represent just the area around our eyes can cause the work done in [7] to converge faster. [7] takes into account the person's face and narrowing this down to just the area around the eye (and including the eye) can further reduce the variables/features associated with this model. This quick and easy solution could further make the computation less expensive by making it converge faster.

We also aim to use the concept of binary thresholding and background subtraction to get the eye to the foreground and use several data augmentation techniques such as flipping the image, tilting the image, resizing, padding, gray-scaling, changing the illumination by modifying the contrast and brightness of the image, saturate the image, sharpening it, blurring it, dilating, performing morphological transformations, etc [11]. We believe doing so will account for the problem that lighting variations and head-pose variations brings to this application.

2. *Training Data*

As part of creating the training dataset, we have 5 calibration points — top left, top right, bottom left, bottom right and the center of the screen. We then use the IR camera to take images of us looking at these points on the screen. We then augment this data using the several image augmentation techniques mentioned above so that it works well with different lighting and head-pose conditions.

3. *Machine Learning*

Here, we design and train a machine learning model using our given dataset. Each input image in the training dataset comes with a vector calculated using the corneal and pupil reflection of the eye in the image [9,10,12,13]. A mapping function that then maps this vector to the 2D screen while calibration can then be used to train the Machine learning model along with the images.

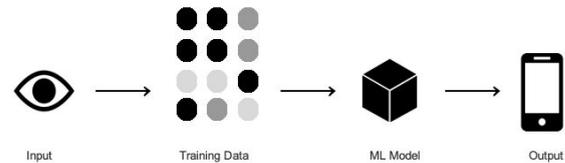

*Figure 2: i-screen Design Diagram*

When the user is presented with this system, the real-time video is broken down into frames per second and these frames are binary thresholded and given to the machine learning model which in turn returns the values of the coordinates of the screen the user is looking at.

## 4. FEASIBILITY ANALYSIS

Our proposed method would be very effective as there will not be any extra modules attached to the user's smartphone, occupying less space. The user will not have to buy modules to have the feature. We aim to have the feature as an additional accessibility feature that uses the smartphone's resources itself.

As mentioned in our solution, we will need smartphones with IR (infra-red) based camera for eye-tracking. Not all phones have them, but phones that use a 3D facial recognition system have the infrared camera built-in. The iPhones use it to take an infrared picture of the user, process it along with other inputs, and authenticate the user. With this, we can expect almost all smartphones to have IR based camera built-in. This solves the requirement of modules that cost in the range from 150-900$.

For the processing power, modern smartphones can run laptop level processes and provide results as fast as a laptop. They are even capable of replacing laptops in certain situations.

With the help of state-of-the-art algorithms, a smartphone's processing power should handle the calibration, training, and mapping process, making it feasible. These cameras also need built-in security features so that the data is protected. Current smartphone manufacturers focus on security a lot more and make sure they are all safe to use.

## 5. CONCLUSION

In this work, we proposed a three-part built-in eye-tracking solution for mobile devices. Firstly, we introduced an additional smartphone setting designed specifically for accessibility called *i-screen*. We then broke down the concrete steps needed for our three-part solution: eye detection, training data, and machine learning. There are several quality applications for tracking eye movement already available, but our proposed solution is a sleek, viable, and useful addition. We hope that with this proposal and future work in accessibility, everyone, including people with ALS, will have full access to their devices.